\documentclass[preprint,showpacs,preprintnumbers,amsmath,amssymb, showkeys]{revtex4}

\usepackage{array}
\usepackage{booktabs}
\usepackage{tabu}
\usepackage{dcolumn}
\usepackage{amsmath}
\usepackage{amsfonts}
\usepackage{amssymb}
\usepackage{graphicx}
\usepackage{subfigure}
\usepackage{graphicx}
\usepackage{dcolumn}
\usepackage{bm}

\begin{document}

\title{Tachyon constant-roll inflation}

\author{A. Mohammadi}
  \email{abolhassanm@gmail.com}
\author{Kh. Saaidi}
  \email{ksaaidi@uok.ac.ir}
  \author{T. Golanbari}
  \email{t.golanbari@gmail.com}

\affiliation{
Department of Physics, Faculty of Science, University of Kurdistan, Sanandaj, Iran.\\}
\date{\today}

\begin{abstract}
The constant-roll inflation is studied where the inflaton is taken as a tachyon field. Since in this approach the second slow-roll parameter is assumed to be of order one instead of being small, then the perturbation parameters will be considered again. The results are compared with observational data, and it is confirmed that the model could stand as a proper candidate for inflation.
\end{abstract}
\maketitle

\section{Introduction}
The Guth's proposal in 1981 for solving the problem of big bang theory is now one of the best candidate for describing the universe evolution in its earliest time \cite{guth}. Although his scenario, known as old inflation, could not work properly, his idea has been followed by physicists and many different scenarios of inflation have been introduced so far such as: new inflation \cite{Linde-a,Albrecht}, chaotic inflation \cite{Linde-b}, k-inflation \cite{Armendariz-Picon,Garriga}, Brane inflation \cite{Maartens,Golanbari}, G-inflation \cite{Abolhasani,Maeda,Alexander,tirandari} warm inflation \cite{berera-a,berera-b,Taylor,Hall,Bastero,sayar,akhtari} and so on. \\
Inflation is known as an accelerated expansion phase in the very early time that the universe undergoes an extreme expansion in a very short period of time. Commonly the inflation is derived by a scalar field, which varies very slowly to give a quasi-de Sitter expansion. Such models of inflationary scenario are known as the slow-roll inflation which describe by slow-roll parameters. In order to give enough amount of expansion, these parameters should be much smaller than unity specially in the early stage of the universe \cite{Linde-c}. The most well-known slow-roll parameters are $\epsilon =-{\dot{H} / H^2}$ which smallness of the parameter means that the fractional change of the Hubble parameter during a Hubble time is much smaller than unity. Usually the same assumption is taken for the time derivative of the scalar field, namely the fractional change of the $\dot\phi$ during a Hubble time (denoting by $\eta = {\ddot{\phi} / H\dot{\phi}}$) should be much smaller than unity \cite{weinberg}. \\
As long as the condition $\epsilon < 1$ is true, the universe remains an accelerated phase, and the smallness of $\eta$ ensure us to have enough inflation in order to overcome the problems of hot big-bang model. It is stated that the smallness of these parameters are provided by almost flat part of the scalar field potential, which in turn results in almost scale invariant spectrum of scalar perturbations. However, what happens if the potential is exactly flat? A question that was explored in \cite{Kinney} for the first time. In \cite{Namjoo}, the non-Gussianity of this situation was explored in detail and they showed that in contrast to the standard slow-roll inflation the non-Gussianity could be of order unity. The idea was generalized in \cite{Martin} so that for the second slow-roll parameter was taken as a constant that could be of order unity. They showed that the power spectrum on superhorizon could be both growing and scale invariant, also the non-Gaussianity of the model could be of order unity too. The model was addressed as ultra slow-roll inflation. The same model was studied in \cite{Motohashi}, where the author found an exact solution for the Hubble parameters and potential using Hamilton-Jacobi formalism \cite{Salopek,Liddle,Kinney,Guo,Aghamohammadi,Saaidi,Sheikhahmadi}. It was determined that the model has an attractor solution, and the power spectrum remains scale invariant for specific choice of constant parameter. Taking into account the higher order terms of curvature in the action, the constant-roll inflation was investigated in modified gravity as well \cite{Motohashi-b,Nojiri,Odintsov-c,Oikonomou-a}. A more generalized case has been considered in \cite{Odintsov-a}, where instead of assuming a constant for $\eta$, it was taken equal to a smooth function of scalar field. Such model is addressed as smooth-roll inflation \cite{Odintsov-a,Odintsov-b,Oikonomou-b}. \\
In this paper, the scenario of constant-roll inflation will be consider for tachyon field. Applications of tachyon field in cosmology was received huge attention after the work of Sen \cite{Sen,Sen-b,Sen-c} and Gibbons \cite{Gibbons}. Sen showed that pressureless gas will be produced during the process of classical decay of unstable D-brane in string theory. The start point of tachyon cosmology was the work of Gibbons \cite{Gibbons}, where the Einstein-Hillbert term was added to the effective action of tachyon on the brane as a way to take into account the coupling to the gravitational field. Since it is assumed that the inflation may be caused by a scalar field (inflaton) with negative pressure and  because of the fact that there is no reason against one to take inflaton as tachyon field, the tachyon inflation becomes an interesting topic amongst physicists \cite{Fairbairn,Mukohyama,Feinstein,Padmanabhan}. This huge interest in tachyon slow-roll inflation motivate us to consider the scenario of tachyon inflation for a more generalized case.      \\

The paper is organised as follow: The general formula of the model is addressed in Sec.II. Applying tachyon field as a inflaton for describing inflation is presented in Sec.III, where the dynamical equations are obtained based on constant-roll condition. In Sec.IV, the scalar and tensor perturbations of the model is extracted and the power spectrum of perturbations are achieved in compatible with constant-roll condition. Agreement of the model prediction with observational data is examined in Sec.V. The result of the work is summarized in Sec.VI.

\section{Tachyon model}
The action for such kind of model is given by
\begin{equation}\label{action}
S = {-1 \over 16 \pi G} \int d^4x \sqrt{-g} R \; + \int d^4x \sqrt{-g} V(T)\sqrt{1 - g^{\mu\nu} \partial_\mu T \partial_\nu T},
\end{equation}
where $T$ stands for tachyon field with potential $V(T)$. The field equation of the model that is derived by variation with respect to the metric is read as
\begin{equation}\label{fieldequation}
R_{\mu\nu} - {1 \over 2} g_{\mu\nu} R = {8\pi G \over 3} \; \left( -V(T)\sqrt{1 +  \partial_\alpha T \partial^\alpha T}
  + {V(T) \over \sqrt{1 +  \partial_\alpha T \partial^\alpha T} }\; \partial_\mu T \partial_\nu T  \right),
\end{equation}
in which the term in parenthesis on the right hand side is known as the energy-momentum tensor of tachyon field. Taking the tachyon field as a perfect fluid with energy-momentum tensor $T_{\mu\nu}= (\rho+p)u_{\mu} u_{\nu} - p g_{\mu\nu}$, one could realized that the four-velocity of the tachyon is $u_{\mu}=\partial_{\mu}T/\big( -  \partial_\alpha T \partial^\alpha T \big)$. The equation of motion of tachyon field is derived by tacking variation of action (\ref{action}) with respect to field $T$, as
\begin{equation}\label{tachyonequation}
\partial_\mu \left( {g^{\mu\nu} \partial_\nu T \; V(T) \over \sqrt{1 - g^{\mu\nu} \partial_\mu T \partial_\nu T}} \right) + V_{,T}(T)\sqrt{1 - g^{\mu\nu} \partial_\mu T \partial_\nu T}=0.
\end{equation}
In homogenous and isotropic universe describing by flat FLRW metric, the main evolution equations of the models are obtained as
\begin{equation}\label{friedmannequation}
H^2 = {8 \pi G \over 3}\; \rho_T, \qquad \dot{H} = -4\pi G \big( \rho_T + p_T \big),
\end{equation}
where $\rho_T$ and $p_T$ are respectively the energy density and pressure of tachyon field given by
\begin{equation}\label{energypressure}
\rho_T = {V(T) \over \sqrt{1-\dot{T}}}, \qquad p_T=-V(T) \sqrt{1-\dot{T}}.
\end{equation}
From Eqs.(\ref{friedmannequation}), the acceleration equation is obtained as $\ddot{a}/a = H^2 (1 - 3\dot{T}^2/2)$, then the universe stay in acceleration expansion phase as long as the condition $\dot{T}^2 < 2/3$, otherwise the universe decelerate. Note that for the case $\dot{T}=0$, we have exactly cosmological constant equation of motion parameter for the scalar field i.e. $\omega=\rho_T / p_T = -1$. \\
The second order equation for tachyon field comes from (\ref{tachyonequation})
\begin{equation}\label{eom}
\ddot{T} + (1-\dot{T}^2)\Big( 3H\dot{T} + {V_{,T}(T) \over V(T)} \Big)=0,
\end{equation}
which is another statement of conservation equation $\dot{\rho}_T + 3H\big( \rho_T + p_T \big)=0$ that also could be concluded form Friedmann equations (\ref{friedmannequation}). If we use the Hamilton-Jacobi formalism and assume the Hubble parameter as a function of tachyon field, namely $H:=H(T)$, from the Friedmann equation (\ref{friedmannequation}), one obtains
\begin{equation}\label{Tdot}
\dot{T} = - {2 \over 3}\; {H_{,T}(T) \over H^2(T)}.
\end{equation}
Substituting Eq.(\ref{Tdot}) in Friedmann equation (\ref{friedmannequation}), and using Eq.(\ref{energypressure}), the potential of tachyon field is derived as
\begin{equation}\label{tachyonpotential}
V(T)= {3 \over 8\pi G} H^2(T)\sqrt{1 - {4 \over 9}\;{H_{,T}^2(T) \over H^4(T)}}.
\end{equation}

\section{Tachyon inflation}
The smallness of the first slow-roll parameter should be guaranteed to have an accelerated phase. Applying the Hamilton-jacobi formalism, and taking the Hubble parameter as a function of scalar field, one arrives at
\begin{equation}\label{firstslp}
\epsilon = - {\dot{H} \over H^2} = - {\dot{T} H_{,T}(T) \over H^2(T)} = {2 \over 3}\; {H_{,T}^2(T) \over H^4(T)}.
\end{equation}

\begin{figure}
  \centering
  \includegraphics[width=7cm]{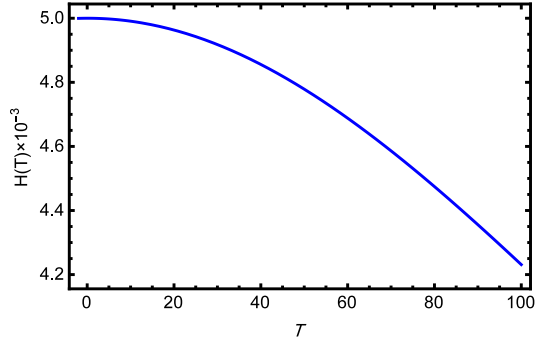}
  \caption{The Hubble parameter versus the tachyon field.}\label{figHubble}
\end{figure}

\begin{figure}
  \centering
  \includegraphics[width=7cm]{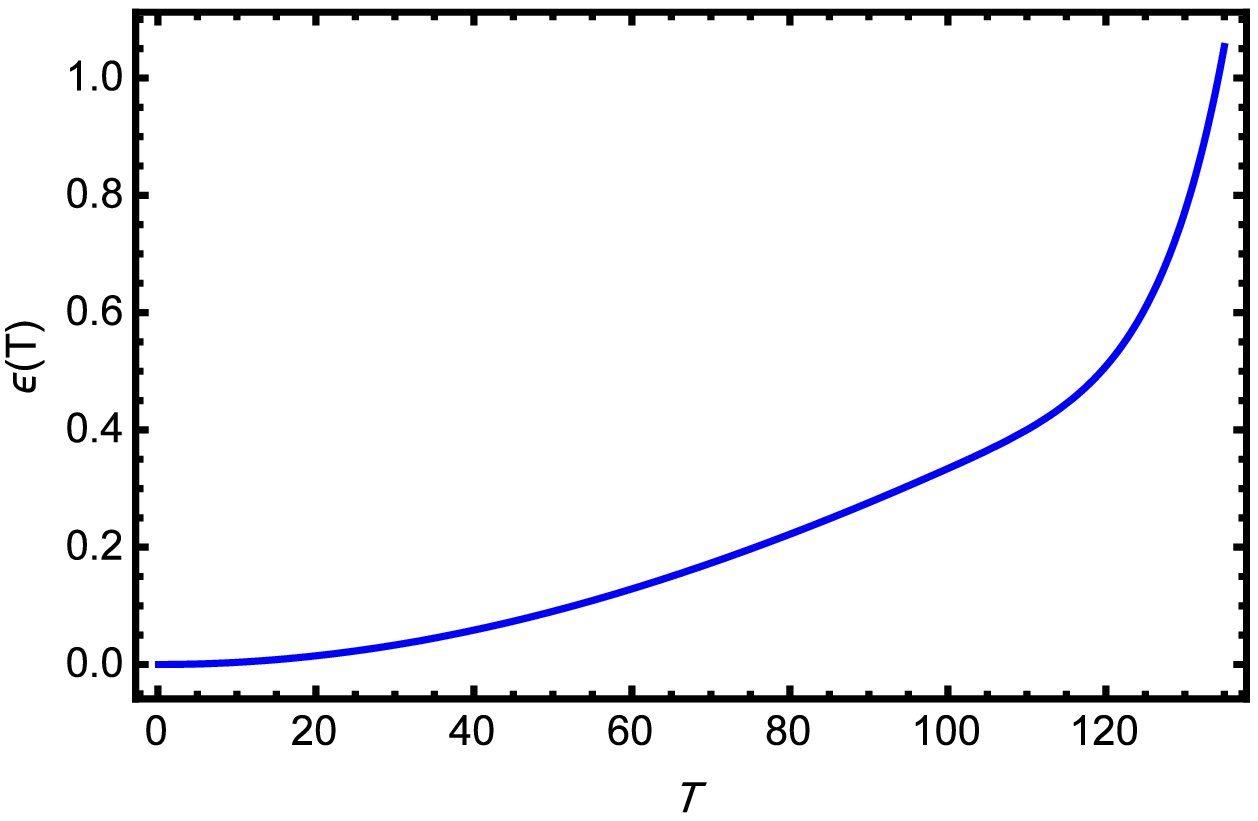}
  \includegraphics[width=7cm]{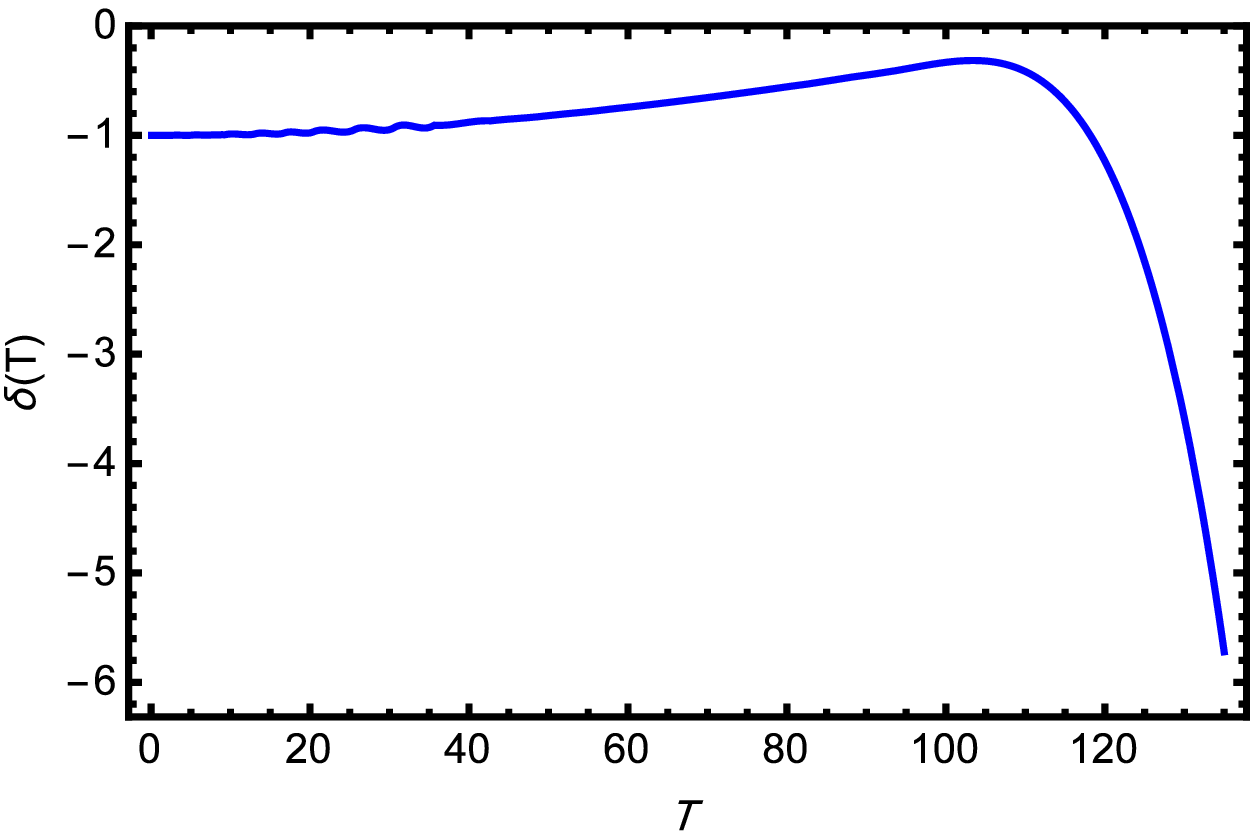}
  \caption{The behavior of the slow-roll parameters $\epsilon$ and $\delta$ during the inflationary times versus tachyon field}\label{figepsilondelta}
\end{figure}

\begin{figure}
  \centering
  \includegraphics[width=7cm]{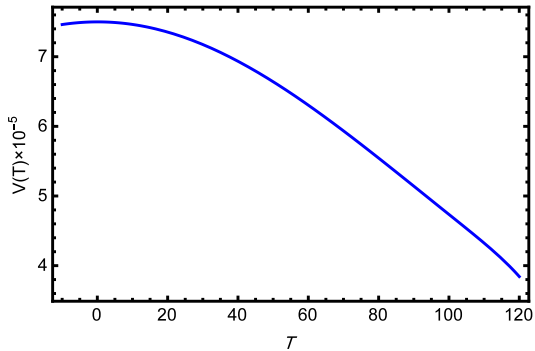}
  \caption{the tachyon field potential versus the field in the inflationary times.}\label{figpotential}
\end{figure}

In constant-roll inflation, the rate of time derivative of field is taken as a constant, namely
\begin{equation}\label{cr}
\ddot{T} = \beta H \dot{T},
\end{equation}
then, our second slow-roll parameter $\eta$ is obtained by
\begin{equation}\label{secondsrp}
\eta = {\ddot{T} \over H\dot{T}} = -\delta + 2\epsilon = \beta, \quad \text{where} \quad \delta= {2 \over 3}\; {H_{,TT}(T) \over H^3(T)}.
\end{equation}
Using Eq.(\ref{Tdot}) and (\ref{cr}), the differential equation for the Hubble parameter is given by
\begin{equation}\label{Hubbleequation}
H(T) {d^2H(T) \over dT^2} - 2\left( dH(T) \over dT \right)^2 + {3 \over 2}\beta H^4(T)=0,
\end{equation}
which is a nonlinear second order differential equation; and it is more complicated than the corresponding equation that in \cite{Motohashi} was derived. Getting analytical solution for the above differential equation is at least very complicated, then we solve it using the numerical approach for specific choice of the parameter $\beta$, illustrated in Fig.\ref{figHubble} where the Hubble parameter decreases by enhancement of the tachyon field. \\
Applying this solution, the behavior of slow-roll parameters $\epsilon$ and $\delta$ will be determined during the inflationary times. Fig.\ref{figepsilondelta}a illustrate the behavior of $\epsilon$ versus tachyon field. The parameter is much smaller than unity at the beginning of inflation and grows up by passing time. It reach one for larger value  of tachyon field. The slow-roll parameter $\delta$ is depicted in Fig.\ref{figepsilondelta}b where one can see that the parameter is close to $\beta$ (not necessary small) at the beginning of inflation, the result that we expected from Eq.(\ref{secondsrp}).\\
From Eq.(\ref{tachyonpotential}), the potential of tachyon field is extracted versus the field. Fig.\ref{figpotential} displays the tachyon potential so that The field rolls down from the top of the potential.\\
Amount of inflationary expansion during this short period of time is important in inflationary studies. The parameter for describing this feature is known as number of e-fold and is given by
\begin{equation}\label{efold}
N = \int_T^{T_e} H dt = -{3 \over 2} \int_T^{T_e} {H^3(T) \over H_{,T}(T)} \; dt
\end{equation}
where $T_e$ indicates the value of the field at the end of inflation. To solve the horizon and flatness problem, we need about $60$ number of e-fold \cite{Baumann}.

\section{Perturbations}
Quantum perturbations are another predictions of inflationary scenario. During the inflationary phase, the fluctuations are stretched and gone beyond the horizon. At superhorizon scale, the fluctuations are frozen and reenter to the horizon after inflation. The footprint of fluctuations are in the cosmic microwave background (CMB) which is emitted freely in the universe at last surface scattering. Therefore, CMB is our best chance to understand the evolution in early time of the universe evolution. Temperature fluctuations in CMB are related to the metric fluctuations, which they in turn are a consequences of scalar field fluctuation (in here tachyon field fluctuations). Metric fluctuations in general are divided to the three types as: scalar, vector, and tensor fluctuations which at linear order they evolve separately. The scalar fluctuations are known as seed for large scale structure of the universe, and tensor fluctuation are addresses as gravitational waves of the early universe. Vector fluctuation are diluted during the inflation and are not our interested perturbations mode. Then, the most important fluctuations types are scalar and tensor fluctuation. In this section, we first consider the scalar fluctuations, and after that introduce the tensor fluctuations. \\

\subsection{scalar perturbation}
Consider an small fluctuation of tachyon field as
\begin{equation*}
T(t,\mathbf{x})= T_0(t) + \delta T(t,\mathbf{x}),
\end{equation*}
where the index $"0"$ indicates that the parameter is a background parameter. The fluctuations in tachyon field cause fluctuations in metric so that the scalar type of these fluctuations in longitudinal gauge is
\begin{equation*}
ds^2 = \big(1 + 2\Phi(t,\mathbf{x}) \big) dt^2 - a^2(t) \big(1 - 2\Phi(t,\mathbf{x})\big)\delta_{ij} dx^i dx^j.
\end{equation*}
Following \cite{Mukhanov,Garriga}, the linearized Einstein equation could be written as
\begin{eqnarray}\label{lineareinstein}
\dot{\xi} & = & {a (\rho_T + p_t) \over H^2} \; \zeta, \nonumber \\
\dot{\zeta} & = & {c_s^2 H^2 \over a^3 (\rho_T + p_t)} \nabla^2 \xi,
\end{eqnarray}
in which the new variables $\xi$ and $\zeta$ are defined as \cite{Garriga}
\begin{eqnarray*}
  \xi &\equiv& {a \over 4\pi G H} \; \Phi, \\
  \zeta &\equiv& {4\pi G H \over a}\; \xi + H {\delta T \over \dot{T}} = \Phi + H {\delta T \over \dot{T}}.
\end{eqnarray*}
The corresponding action for the equation (\ref{lineareinstein}) is \cite{Garriga}
\begin{equation}\label{zetaaction}
  S = {1 \over 2} \int z^2 \left( \zeta'^2 + c_s^2 \zeta (\nabla\zeta)^2 \right) d\tau d^3\mathbf{x},
\end{equation}
and the parameter $z$ is given by $z=\sqrt{3/8\pi G}\; \big( a \dot{T} / c_s \big)$. The prime in above equation means derivative with respect to the conformal time $\tau$. Introducing a canonical quantization variable $v=z\zeta$, the action (\ref{zetaaction}) is rewritten in terms of $v$, , which is known as Mukhanov-Sasaki variable
\begin{equation}\label{vaction}
  S = {1 \over 2} \int  \left( v'^2 + c_s^2 v (\nabla v)^2 + {z'' \over z} v \right) d\tau d^3\mathbf{x}.
\end{equation}
Therefore, we have
\begin{equation}\label{vequation}
{d^2 \over d\tau^2} v(\tau,\mathbf{x}) - c_s^2 \nabla^2 v(\tau,\mathbf{x}) - {z'' \over z} \; v(\tau,\mathbf{x})=0.
\end{equation}
The conformal time derivatives of $z$ is extracted as
\begin{equation*}
  {dz \over d\tau} = z (aH)\left[ 1 + 2\epsilon - \delta + {2\epsilon \over (3 - 2\epsilon)}\; \big( 2\epsilon - \delta \big) \right].
\end{equation*}
In order to get the second time derivative, one need to calculate the time derivative of the slow-roll parameters
\begin{eqnarray*}
  {d\epsilon \over d\tau} &=& aH \big( 2\epsilon\delta - 4\epsilon^2 \big), \\
  {d\delta \over d\tau} &=& aH \big( \sigma^2 -3\epsilon\delta \big),
\end{eqnarray*}
where $\sigma= 2\sqrt{H' H'''} / 3H^3$. Then, the second order derivative of $z$ with respect to the conformal time becomes
\begin{eqnarray*}
   {1 \over z}\;{d^2z \over d\tau^2} & = & 2 \big( aH \big)^2 \left[ \Big(1+ {5 \over 2}\epsilon - {3 \over 2} \delta - 3\epsilon^2
                                               + {1 \over 2}\delta^2 + 2\epsilon\delta - {1 \over 2} \sigma^2 \Big) \right. \\
          &   & \hspace{2cm} +{\epsilon \over (3 - 2\epsilon)}
                            \Big( 4\epsilon - 2\delta - 8\epsilon^2 + 7\epsilon\delta - \sigma^2 \Big) \\
          &   &\hspace{2cm} + \left. {2\epsilon^2 \over 3(3 - 2\epsilon)^2}
                                           \Big( -4\epsilon^2 + 4\epsilon\delta - \delta^2 \Big) \right].
\end{eqnarray*}
Now, we back to the main equation (\ref{vequation}). By expanding the canonical quantization variable $v$ in the Fourier modes as
\begin{equation*}
  v(\tau,\mathbf{x}) = \int {d^3\mathbf{k} \over \big( 2\pi \big)^{3/2}} \; v_k(\tau) e^{i\mathbf{k}.\mathbf{x}},
\end{equation*}
and inserting this in Eq.(\ref{vequation}), we arrive at \cite{Mukhanov}
\begin{equation}\label{vkequation}
{d^2 \over d\tau^2} v_k(\tau) +\left( c_s^2 k^2 - {1 \over z}\;{d^2z \over d\tau^2} \right) \; v_k(\tau)=0.
\end{equation}
We need to find $v_k(\tau)$ from above equation to extract the power spectrum. The power spectrum of curvature perturbations as a function of wavenumber $k$ is given by
\begin{equation}\label{PS1}
\mathcal{P}_s = {k^3 \over 2\pi^2} \; \left| {v_k \over z} \right|^2.
\end{equation}
At subhorizon scale, when $c_s k \gg aH$, the differential equation (\ref{vkequation}) gets a simple form which the solution could be easily derived \cite{Mukhanov,Garriga}
\begin{equation}\label{subv}
  {d^2 \over d\tau^2} v_k(\tau) + c_s^2 k^2  v_k(\tau)=0, \quad \Rightarrow \quad
             v_k(\tau)={1 \over \sqrt{2c_s k}} e^{ic_s k \tau}.
\end{equation}
To find a general solution, we first work a little with term $z''/z$. Since the slow-roll parameters appeared in $z''/z$ are smaller than unity, except $\eta$ (or $\delta$), we rewrite this term up to first order of $\epsilon$ and $\sigma$, namely
\begin{equation*}
   {1 \over z}\;{d^2z \over d\tau^2} = 2 \big( aH \big)^2 \Big(1+ {5 \over 2}\epsilon - {3 \over 2} \delta
                                               + {1 \over 2}\delta^2 + {4 \over 3} \epsilon\delta  \Big).
\end{equation*}
On the other side, for a quasi-de Sitter expansion (which is at least acceptable for the beginning of inflation) the conformal time is derived in terms of the scale factor and Hubble parameter as $\tau=-(1+\epsilon)/aH$, therefore, there is $a^2H^2 \simeq (1+\epsilon)^2/\tau^2$. Substituting this in term $z''/z$, we could have
\begin{equation*}
   {1 \over z}\;{d^2z \over d\tau^2} = {\nu^2 - {1\over 4} \over \tau^2}, \qquad
                      \nu^2 \equiv {9 \over 4} + 9 \epsilon - 3 \delta + \delta^2 - {10 \over 3} \epsilon\delta + 2\delta^2\epsilon.
\end{equation*}
Inserting this in Eq.(\ref{vkequation}), and using a variable change as $v_k=\sqrt{-\tau} \; f_k$, the canonical form of Bessel's equation is extracted so that
\begin{equation}\label{bessel}
  {d^2 f_k \over dx^2} + {1 \over x}{d f_k \over dx} + \Big( 1 - {\nu^2 \over x^2} \Big)f_k=0.
\end{equation}
The most general solution for the equation is the Hankel functions. Therefore, we have
\begin{equation}\label{vhankel}
v_k(\tau) = \sqrt{-\tau} \; \left[ \alpha_k H_\nu^{(1)}(x) + \gamma_k H_\nu^{(2)}(x) \right],
\end{equation}
where $\alpha_k$ and $\gamma_k$ are constant that generally depend on wavenumber $k$. The solution should approach to solution (\ref{subv}) on subhorizon scale (as $-k\tau \rightarrow \infty$). The asymptotic behavior of the Hankel functions at this limit is given by
\begin{equation*}
  \lim_{-k\tau \rightarrow \infty}H_\nu^{(1,2)}(x) = \sqrt{{2 \over \pi}} \; {1 \over \sqrt{2c_s k}} e^{\mp (ic_s k\tau + \varsigma)},
         \qquad  \varsigma={1 \over 2} \big( \nu + {1 \over 2} \big).
\end{equation*}
In comparison with solution (\ref{subv}), it is clear that $\alpha_k = \sqrt{\pi} \; e^{i\delta} / 2$ and $\gamma_k=0$. Inserting this into Eq.(\ref{vhankel}), the general solution for variable $v_k(\tau)$ becomes
\begin{equation}\label{vsolution}
v_k(\tau) = {\sqrt{\pi} \over 2} e^{i{\pi \over 2} (\nu+1/2)} \sqrt{-\tau} H_\nu^{(1)}(-c_sk\tau).
\end{equation}
The scalar power-spectrum follow from Eq.(\ref{PS1}), and is read as
\begin{equation}\label{spectrum}
\mathcal{P}_s = {1 \over 2\pi}\; {8\pi G \over 3}\; {3c_s^2 H^2 \over 2\epsilon}\; \left({k \over aH}\right)^3
                \left| H_\nu^{(1)}(-c_sk\tau) \right|^2.
\end{equation}
Following the asymptotic behavior of the Hankel function, the scalar power spectrum on superhorizon scale becomes
\begin{equation}\label{PSsuperhorizon}
\mathcal{P}_s = {8\pi G \over 3}\; \left( {H \over 2\pi} \right)^2 {3 \over 2 c_s \epsilon}\;
                \left( {2^{\nu-3/2} \Gamma(\nu) \over \Gamma(3/2)} \right)^2 \; \left( {c_s k \over aH} \right)^{3-2\nu}.
\end{equation}
Then the scalar spectral index is $n_s-1=3-2\nu$.

\subsection{Tensor perturbations}
Besides scalar perturbations, quantum fluctuations in gravitational waves form are produced as the same way. A linear perturbations in flat FLRW metric is given by
\begin{equation*}
ds^2 = -dt^2 + \big( \delta_{ij}+h_{ij} \big) a^2(t) dx^i dx^j,
\end{equation*}
the perturbation quantity $h_{ij}$ is gauge invariant. Since the energy-momentum tensor is diagonal, therefore it has no contribution in tensor perturbations equations; in another word the tensor mode equation has no source term. The action for the corresponding equation is given by \cite{Baumann,Riotto}
\begin{equation}\label{tensoraction}
  S = {1 \over 2G} \; \int d^4x \sqrt{-g} {1 \over 2} \partial_\mu h_{ij} \partial^\mu h^{ij},
\end{equation}
which has the same form as the action of massless scalar field. Utilizing $u_k=ah_k / \sqrt{2} G$, and repeating the same process, the following equation will be concluded
\begin{equation}\label{uequation}
{d^2 u_k(\tau) \over d\tau^2} + \Big( k^2 - {1\over a}{d^2 a \over d\tau^2} \Big)u_k(\tau)=0,
\end{equation}
which the solution is now familiar. On superhorizon limit, it is read as
\begin{equation}\label{usuperhorizon}
\left| u_k \right| = \left( H \over 2\pi \right) \left( {k \over aH} \right)^{-\nu_T + 3/2}, \qquad \nu_T = {3 \over 2} - \epsilon.
\end{equation}
In turn, the tensor power spectrum at this limit is obtained as
\begin{equation}\label{Pt}
\mathcal{P}_T = 64 \pi G \; \left( {H \over 2\pi} \right)^2
                 \left( {k \over aH} \right)^{n_T},
\end{equation}
and the tensor spectral index is $n_T=3-2\nu_T$. The tensor perturbations are detected indirectly through measuring the tensor-to-scalar ratio parameter $r = \mathcal{P}_T / \mathcal{P}_s$. Although no exact value for the parameter has been measured, there is only an upper bound so far.

\section{Consistency with observation}
\begin{figure}[h]
  \centering
  \includegraphics[width=7cm]{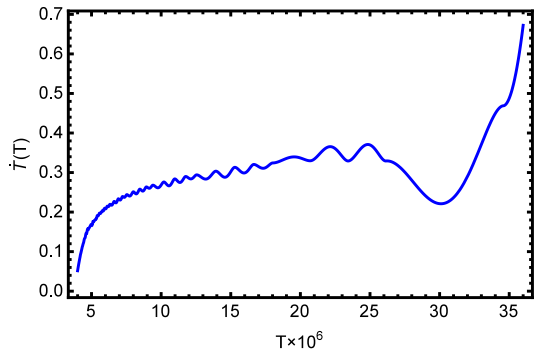}
  \caption{The time derivative of the tachyon field versus the field is plotted during the inflation. }\label{phidotfig}
\end{figure}

\begin{figure}[h]
  \centering
  \includegraphics[width=7cm]{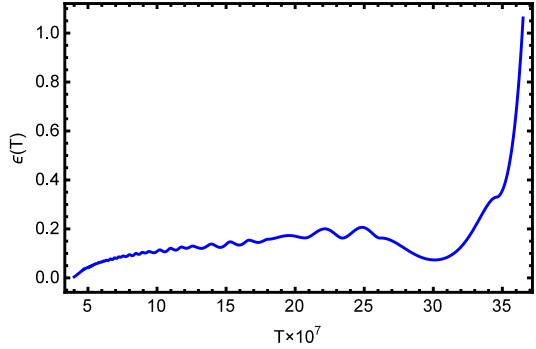}
  \caption{The slow-roll parameters $\epsilon$ versus tachyon field during inflation. }\label{epsilonfig}
\end{figure}

\begin{figure}[h]
  \centering
  \includegraphics[width=7cm]{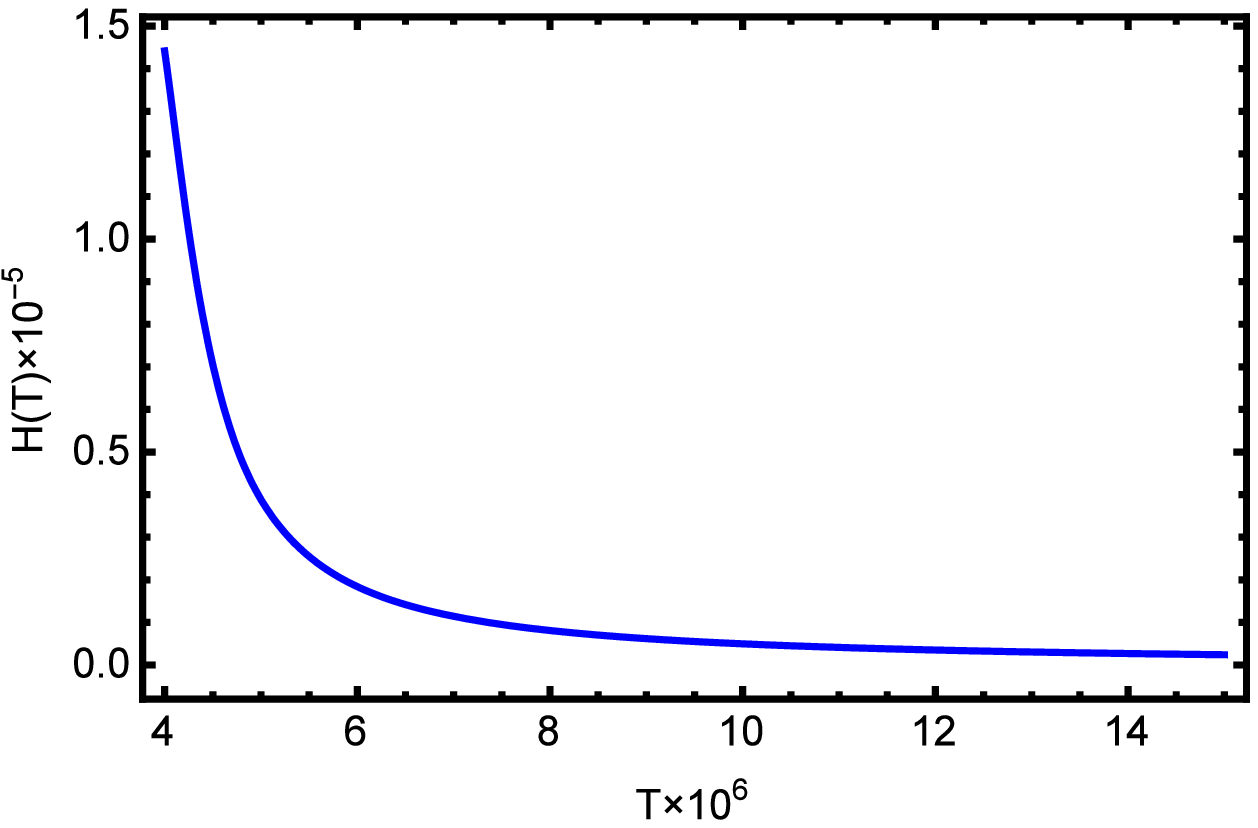}
  \includegraphics[width=7cm]{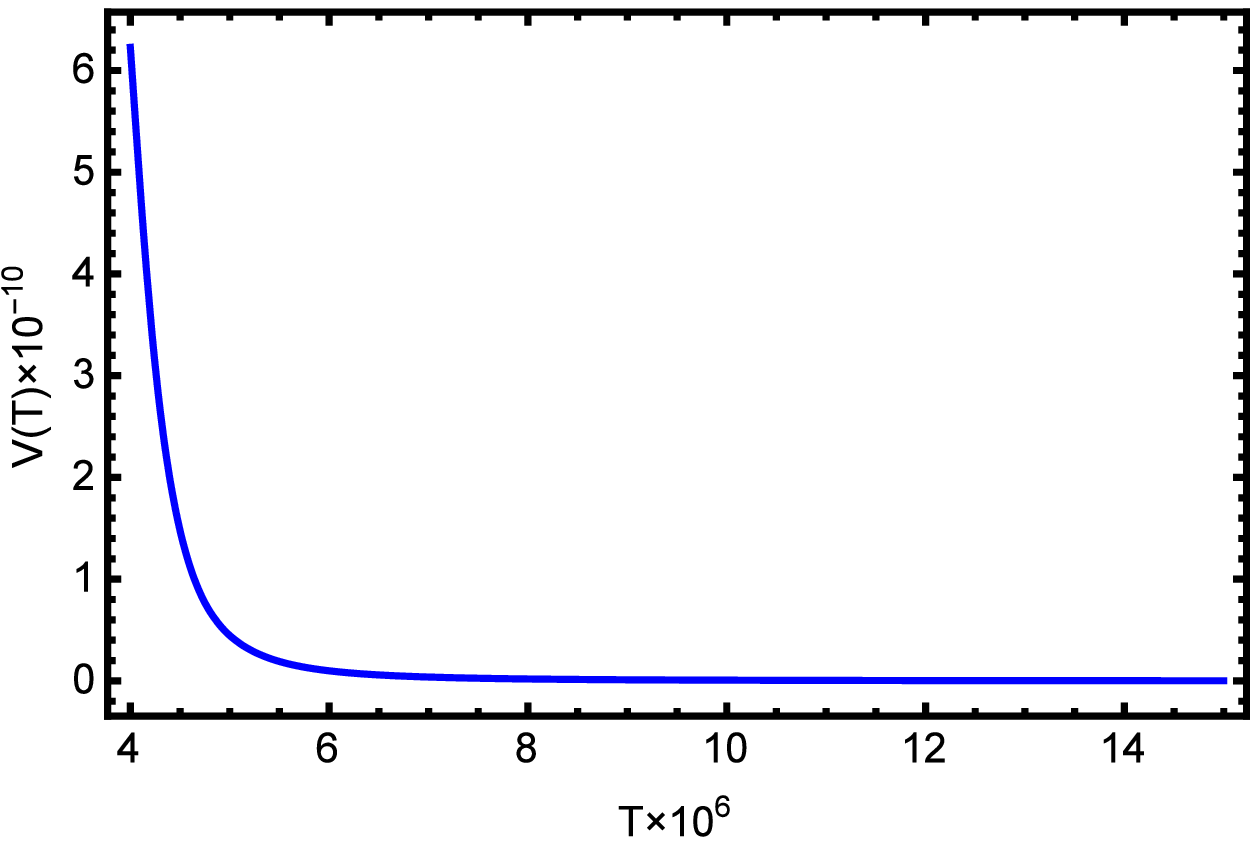}
  \caption{a) The Hubble parameter and b) the potential versus tachyon field during inflation. }\label{HPfig}
\end{figure}

The model was studied generally, however to get the best solution and in order to confirm the validity of the model, its prediction should be compare with observational data. The scalar spectral index in terms of $\eta$ is read by
\begin{equation}\label{ns}
n_s - 1 = 3 - 2\nu = -2\eta, \qquad
\nu^2 = {9 \over 4} + 3 \epsilon + 3 \eta + \eta^2 - {2 \over 3} \epsilon\eta + 2\eta^2\epsilon
\end{equation}
in which $\eta=\beta$. According to the Planck observational data, the scalar spectral index is about $n_s=0.9666 \pm 0.0033$ $68\%$ CL \cite{Planck}. Eq.(\ref{ns}) indicates that there are two values for $\beta$ as $\beta \approx 0.0153$ and $\beta \approx -2.989$ to achieve the proper value for $n_s$. For $\beta \approx 0.0153$, the time derivative of tachyon field is positive. Fig.\ref{phidotfig} the behavior of $\dot{T}$ versus the tachyon field illustrating that $\dot{T}>0$ during the inflation, denoting that the tachyon field at the end of inflation is bigger than it at the horizon crossing.
The slow-roll parameter $\epsilon$ is depicted in Fig.\ref{epsilonfig}, where it is clear that for small values of the field $\epsilon$ is smaller than unity correspond to the horizon crossing time. However, for larger values of the field, $\epsilon$ reaches one and inflation ends.
The Hubble parameters and also the potential of tachyon field is shown in Fig.\ref{HPfig} so that the potential rolls down from the top of the potential.
For this choice of the parameter $\beta$ there are the scalar spectral index $n_s \approx 0.9615$, the amplitude of scalar perturbations $\mathcal{P}_s \approx 2.13 \times 10^{-9}$, and the tensor-to-scalar ratio $r \approx 0.0197$, however these results are obtained for about $N=121$ number of e-folds.

\section{Very smooth-roll tachyon inflation}
\begin{figure}[h]
  \centering
  \includegraphics[width=7cm]{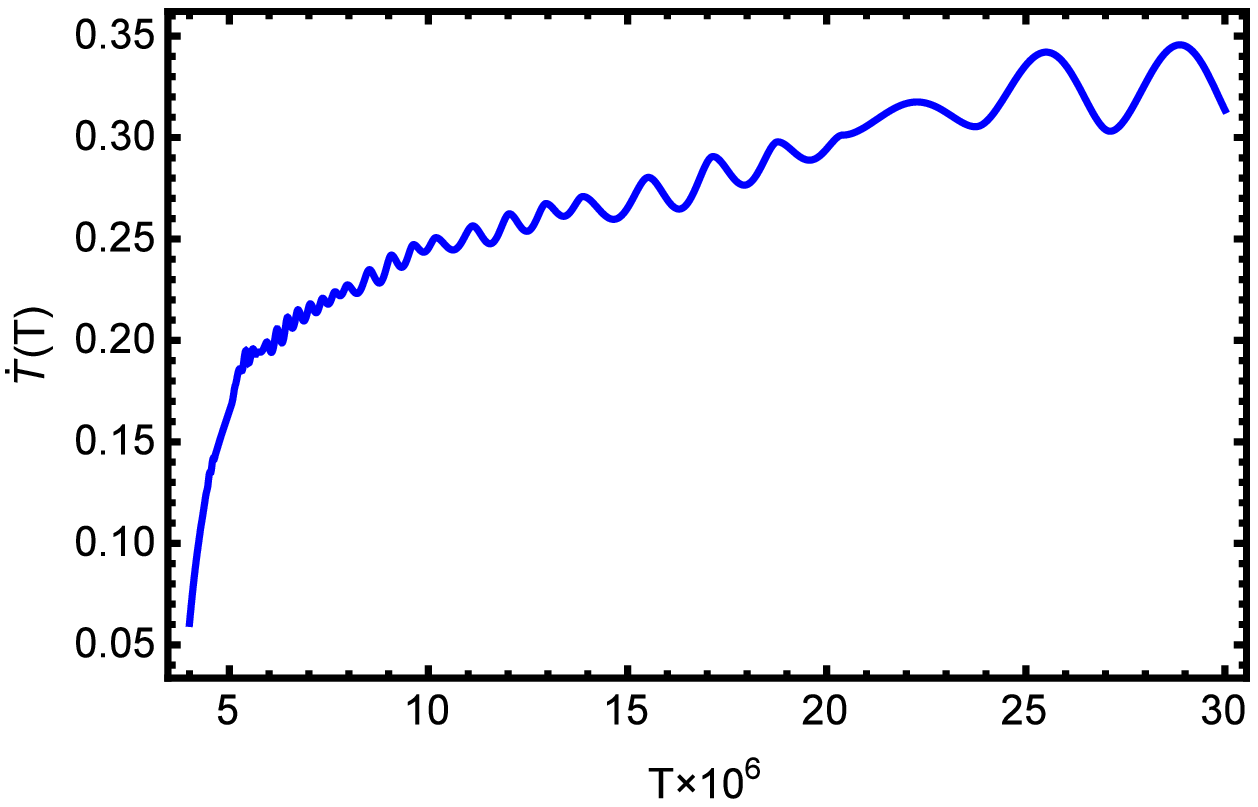}
  \includegraphics[width=7cm]{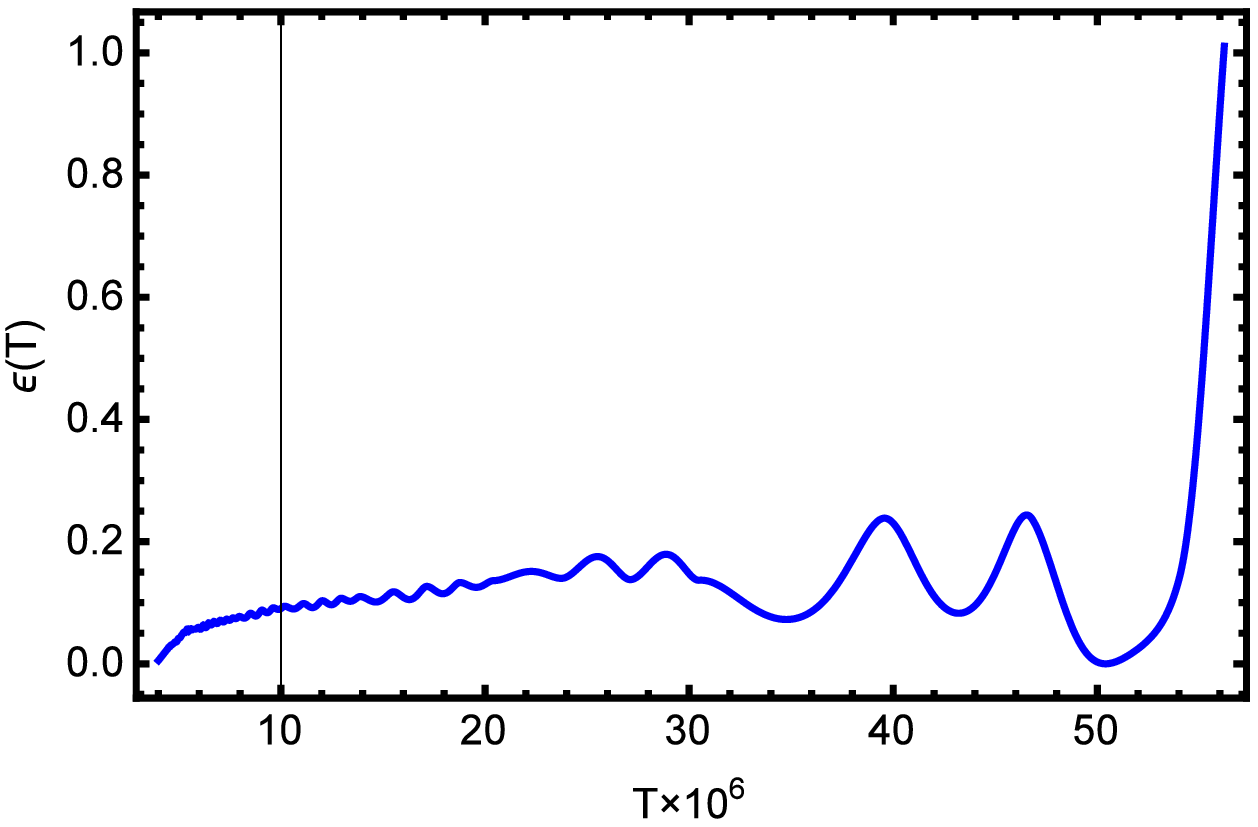}
  \caption{a) The time derivative of tachyon field, and b) the behavior of the slow-roll parameter $\epsilon$ are plotted versus the field during the inflation. }\label{Sphidotepsilonfig}
\end{figure}

\begin{figure}[h]
  \centering
  \includegraphics[width=7cm]{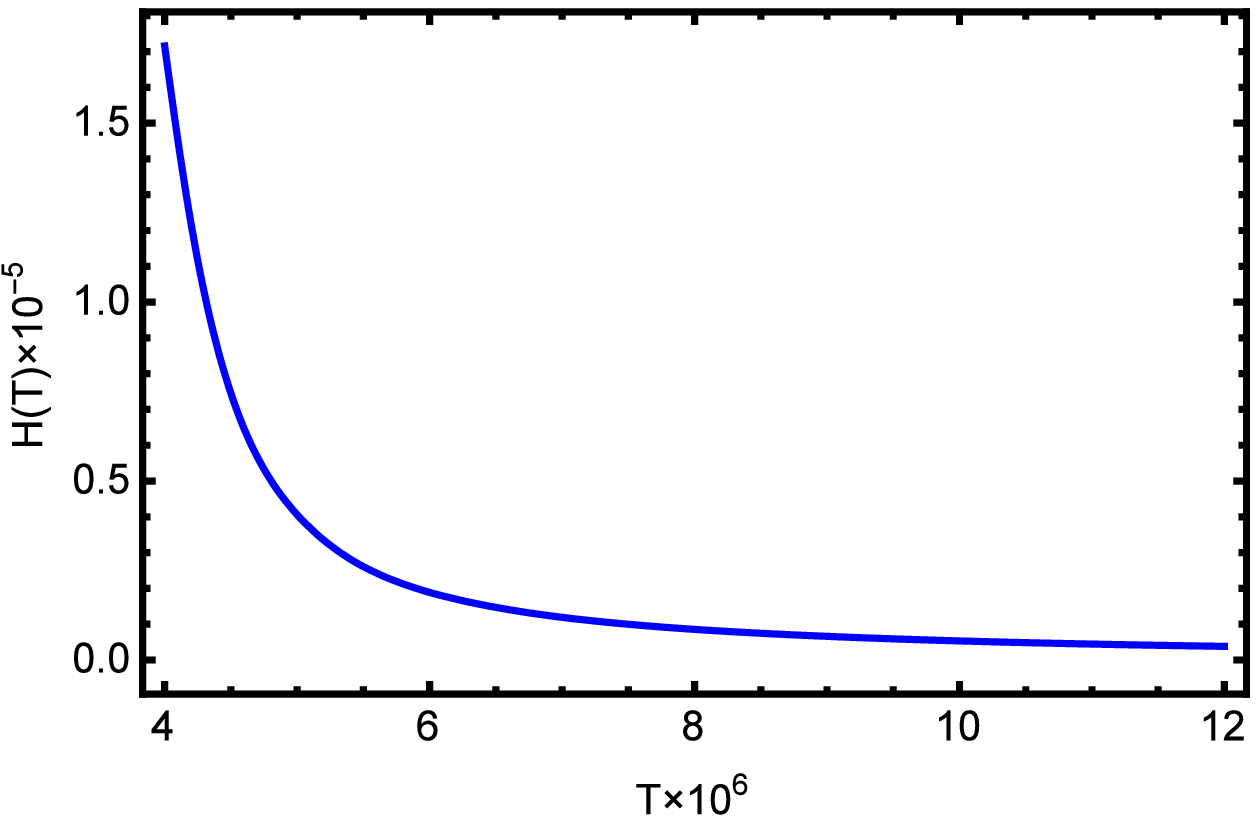}
  \includegraphics[width=7cm]{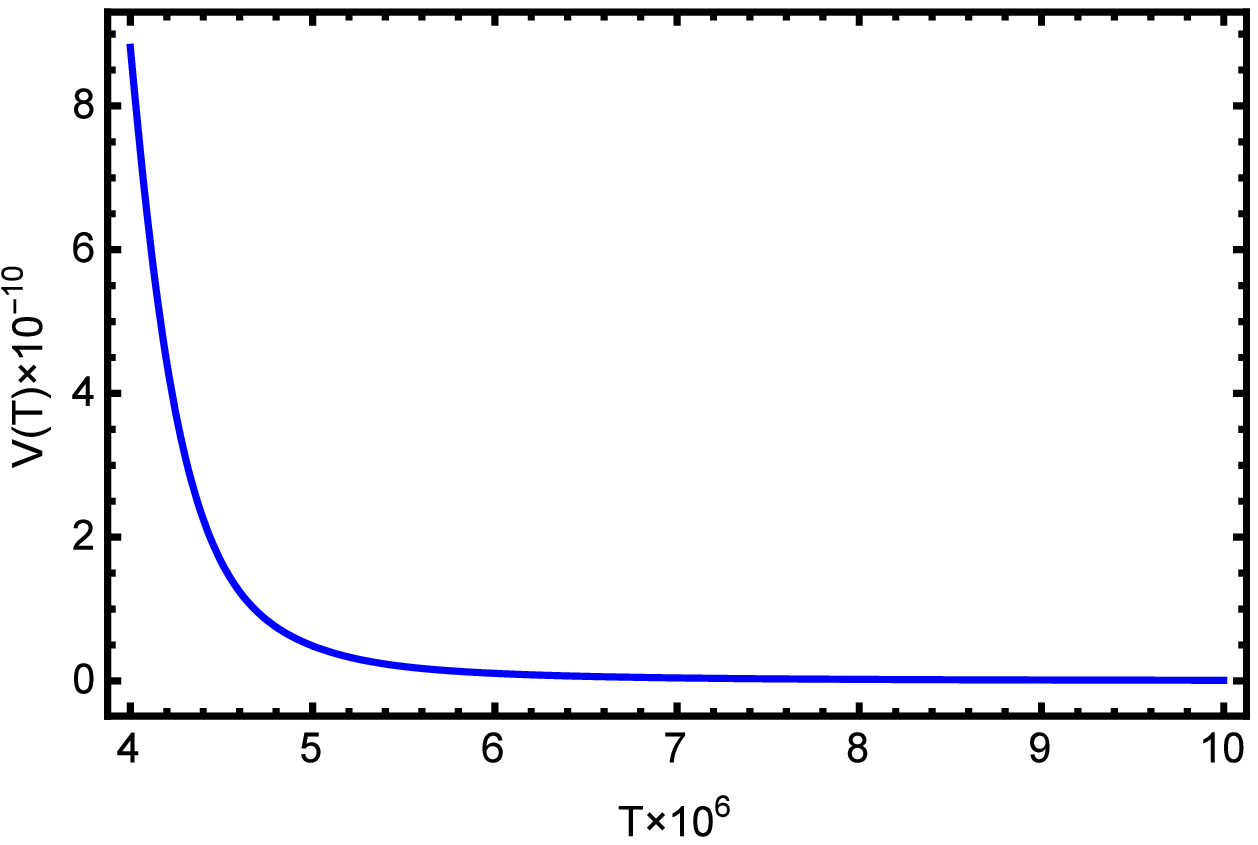}
  \caption{a) the Hubble parameter, and b) the potential of scalar field are plotted versus the field during the
  inflation. }\label{Shubblepotfig}
\end{figure}

In the canonical scalar field model of constant-roll inflation, for $\beta=-3$ there is a flat potential i.e. $V'(\phi)=0$ \cite{Martin,Motohashi}. However, in the tachyon inflation, this choice does not lead to a flat potential; as a matter of fact for $\beta = -3$ we have
\begin{equation*}
  {V'(T) \over V(T)} = {3H\dot{T}^3 \over 1-\dot{T}^2 }
\end{equation*}
and this is due to the different form equation of motion of tachyon field. On the other side, for $\eta = \beta \big( 1 - \dot{T}^2 \big)$, and $\beta=-3$, a flat potential is concluded. This choice for $\eta$ is not constant, actually it is varies slowly and for this reason we call this choice "very smooth-roll", because it could be written as $\eta = \beta (1-2\epsilon/3)$. \\
In this case, the differential equation for the Hubble parameters changes a little and is
\begin{equation}\label{Hubbleequationb}
H(T) {d^2H(T) \over dT^2} - 2\Big( 1 + {\beta \over 3} \Big)\left( dH(T) \over dT \right)^2 + {3 \over 2}\beta H^4(T)=0,
\end{equation}
however, all of the results we have obtained in Sec.III will be right even for this choice of $\eta$, and only there is a difference in the parameter $\nu$ as
\begin{equation*}
  \nu^2 = {9 \over 4} + 9 \epsilon - 3 \beta - 4\beta\epsilon + \beta^2 + 2\beta^2 \epsilon,
\end{equation*}
which in turns appears in the scalar spectral index so that up to the first order of $\epsilon$ there is
\begin{equation}\label{nssmooth}
n_s -1 =  3 - 2\nu.
\end{equation}
It seems that for any value of $n_s$ there are two values for $\beta$. For $\beta \approx 0.0124$, the time derivative of the tachyon field is positive, illustrating in Fig.\ref{Sphidotepsilonfig}a. The slow-roll parameter $\epsilon$ is smaller than unity at the horizon crossing time. By passing time, the tachyon field increases and $\epsilon$ also approached one. Its behavior is shown in Fig.\ref{Sphidotepsilonfig}b and it is seems that $\epsilon$ reaches one for bigger values of field than the previous case. \\
The Hubble parameter and the potential of tachyon field for this case are also displayed respectively in Fig.\ref{Shubblepotfig}a and Fig.\ref{Shubblepotfig}b which in general have the same behavior at their corresponding in the previous case. \\
For this choice of $\beta$, the scalar spectral index is about $n_s \approx 0.9645$, the amplitude of scalar perturbations is obtained about $\mathcal{P}_s \approx 2.19 \times 10^{-9}$, and tensor-to-scalar ratio is predicted about $r \approx 0.0272$, for about $N=108$ number of e-folds.

\section{Conclusion}
Tachyon inflation was studied in constant-roll scenario where the second order of the slow-roll parameter is assumed as a constant. This choice leads to a nonlinear differential equation for the Hubble parameter, and it complicity form forced us to solve it numerically.     \\
Since this constant taken for the second slow-roll parameter could be of order unity, one should reinvestigate the cosmological perturbations of the model. The cosmological perturbation was considered in the work, and for cosmological perturbation it was clear that there are some correction terms in the amplitude of scalar perturbations and also in the scalar spectral index. However, since the energy-momentum tensor has no contribution in the tensor perturbations equations, and also because these equations only contain the first slow-roll parameter, there was no change in the amplitude of tensor perturbation with respect to the slow-roll inflationary model. \\
Obtaining the perturbations parameters and comparing them with observational data, indicates that there is a specific choice for the constant parameter $\beta$ that makes the amplitude of scalar perturbation almost scale invariant in superhorizon scales. Taking $\beta \approx 0.024$, comes to a positive time derivative of tachyon field which shows that the field grows up during the inflationary times, that is consistent with Fig.\ref{epsilonfig} where the behavior of the slow-roll parameters $\epsilon$ is plotted and clearly shows that inflation ends for bigger value of the field. The amplitude of scalar perturbations, scalar spectral index, and tensor-to-scalar ratio for the selected $\beta$ are respectively obtained as $0.9615$, $2.13 \times 10^{-9}$, and $0.0197$ for about $121$ number of e-folds.\\
As another case, the second slow-roll parameter was taken as $\eta=\beta\big(1-\dot{T}^2\big)$ so that for $\beta=-3$ there is flat potential for the model in analogues with the canonical scalar field model that we have a flat potential for $\eta=-3$. In this case, the parameter $\eta$ is not exactly constant, and it varies slowly. The differential equation for the Hubble parameter changes a little with respect to the previous case. Calculating the perturbation equations shows that there is only some correction terms in the parameter $\nu$, that in turns comes to some correction terms in the equation of the scalar spectral index. By taking $\beta \approx 0.0233$, the time derivative of the field is positive indicating an increasing behavior for the magnitude of the field, which displays that there is an end for inflation for the bigger values of the field than horizon crossing time, illustrating in Fig.\ref{Sphidotepsilonfig}b. The results states that there are $n_s \approx 0.9645$, $\mathcal{P}_s \approx 2.19 \times 10^{-9}$, and $r \approx 0.0272$ with about $108$ number of e-folds.





\end{document}